\documentclass[conference]{IEEEtran}
\IEEEoverridecommandlockouts
\usepackage{cite}
\usepackage{amssymb,mathtools}
\usepackage{graphicx, times, amsmath, amsfonts, comment,amsthm}
\usepackage{algorithmic}
\usepackage{algorithm, array}

\ifCLASSOPTIONcompsoc
    \usepackage[caption=false, font=normalsize, labelfont=sf, textfont=sf]{subfig}
\else
\usepackage[caption=false, font=footnotesize]{subfig}
\fi

\usepackage{epstopdf}
\usepackage{tikz}
\begin{document}

\title{A Deep Q-Learning Method for Downlink Power Allocation in Multi-Cell Networks}
%
%
%
\author{Kazi Ishfaq Ahmed and Ekram Hossain
\thanks{K. I. Ahmed and E. Hossain are with the Department of Electrical and Computer Engineering at the University of Manitoba, Canada. 
(Email:  ahmedki@myumanitoba.ca, Ekram.Hossain@umanitoba.ca). 
}}

\markboth{Journal of \LaTeX\ Class Files,~Vol.~14, No.~8, August~2015}%
{Shell \MakeLowercase{\textit{et al.}}: Bare Demo of IEEEtran.cls for IEEE Journals}

\maketitle

\begin{abstract}
Optimal resource allocation is a fundamental challenge for dense and heterogeneous wireless networks with massive wireless connections. Because of the non-convex nature of the optimization problem, it is computationally demanding to obtain the optimal resource allocation. Recently, deep reinforcement learning (DRL) has emerged as a promising technique in solving non-convex optimization problems. Unlike deep learning (DL), DRL does not require any optimal/ near-optimal training dataset which is either unavailable or computationally expensive in generating synthetic data. In this paper, we propose a novel centralized DRL based downlink power allocation scheme for a multi-cell system intending to maximize the total network throughput. Specifically, we apply a deep Q-learning (DQL) approach to achieve near-optimal power allocation policy. For benchmarking the proposed approach,  we use a Genetic Algorithm (GA) to obtain near-optimal power allocation solution. Simulation results show that the proposed DRL-based power allocation scheme performs better compared to the conventional power allocation schemes in a multi-cell scenario.   
\end{abstract}


\IEEEpeerreviewmaketitle

\section{Introduction}

Optimal resource allocation will be a crucial problem in future 6G wireless communications \cite{david20186g} because of the massive connections and the ultra-dense deployment of base stations on a large scale. Traditionally, due to the non-convex nature of the optimization problem, resource allocation is done using some heuristic approaches such as exhaustive search, genetic algorithms, combinatorial and branch and bound techniques. These methods are computationally expensive and therefore not feasible for large-scale cellular networks. Recently, supervised deep learning (DL) \cite{goodfellow2016deep}-based resource allocation schemes \cite{ahmed2018deep,zappone2018online,sanguinetti2018deep} are proposed where the training data is generated through some heuristic algorithms such as GA, sequential fractional programming (SFP), bisection approach, etc. The training data generation is computationally expensive and time-consuming, and therefore, the supervised DL approach is not suitable for large-scale systems. 

On the other hand, Reinforcement Learning (RL) method can obtain the optimal solution of a control problem by interacting with the environment. Q-learning \cite{watkins1992q} is a widely used reinforcement learning and is already applied for cognitive radio applications \cite{bennis2010q,li2010multiagent}.  An agent in Q-learning interacts with the environment by taking an action and then receiving feedback from the environment in terms of reward. The agent follows a policy to maximize some notion of discounted cumulative reward through a series of actions. Occasionally, the environment in RL is formulated as a Markov Decision Process (MDP). The agent updates policy depending on the rewards from the environment. Through several interactions with the environment, the agent learns the optimal action policy. However, the traditional RL is suitable only for systems with  low-dimensional state space and may not work for systems with high-dimensional state space. This is because, in traditional RL, a policy is stored in tabular form and it is not feasible for large action and state space due to the lack of generalization. So, instead of a tabular method, a function approximation such as a deep neural network (DNN) \cite{goodfellow2016deep} can be used in that case. Recently, DRL \cite{mnih2015human} has emerged as a promising technique to handle complicated control problems. By combining Deep Learning (DL) with Reinforcement Learning (RL), the DRL can extract useful information from high-dimensional data and can learn the optimal action policy. 

Power allocation under maximal power constraints in a multi-cell network (e.g. a cloud-RAN) to maximize the total network throughput is a well-known non-convex combinatorial optimization problem and is NP-hard \cite{luo2008dynamic}. Model-based algorithms such as Fractional Programming (FP) \cite{shen2018fractional} and Weighted Minimum Mean Squared Error (WMMSE) \cite{shi2011iteratively} are usually used in this scenario. However, both the algorithms formulate the power allocation problem as a convex optimization problem. 

In the above context, the contributions of this paper can be summarized as follows:
\begin{itemize}
    \item A centralized downlink power allocation scheme based on DRL is proposed for multi-cell network  to maximize the total network throughput. Our proposed scheme is novel and it is one of the first such schemes to address the power allocation problem under maximal power constraints in a multi-cell network having multiple users sharing the same frequency subbands. 
    \item We define the state space, action space and the reward function for the DRL agent. We also define the online training procedure of the proposed DRL based power allocation scheme. 
    \item Unlike supervised learning approach, there is no need for optimal/ near-optimal training data. This is why our proposed power allocation scheme is computationally scalable to large-scale scenarios. 
    \item Simulation results with different network size and training parameters are presented to show the scalability and robustness of our algorithm. We also compare our DRL model with the near-optimal solution derived through a GA. Simulation results show that our model can perform well in a large-scale scenario. 
\end{itemize}

The rest of the paper is organized as follows. In Section \ref{literature}, a brief review of the existing DRL-based resource allocation schemes is presented. In Section \ref{system}, we introduce the system and formulate the power allocation problem. 
In Section \ref{proposed}, we present our DRL-based power allocation scheme in detail. In Section \ref{result}, we discuss the simulation results followed by the conclusion in Section \ref{concl}.    

\section{Related Work} \label{literature}
DRL techniques have recently been used in a variety of wireless resource management problems (e.g. channel and power allocation, throughput maximization, spectrum access). 
In \cite{xu2017deep}, the authors use a DQL for power allocation in a cloud-RAN to minimize the total power consumption while ensuring the demand of each user. 
In \cite{li2018intelligent}, the authors develop a distributed DQL based spectrum sharing approach for primary and secondary users in a non-cooperative fashion. The primary users use a fixed power control strategy while the secondary users learn autonomously to adjust the transmission power to share the shared spectrum. 
In \cite{zhao2018deep}, the authors use the DRL approach to perform joint user association and resource allocation (UARA) in the downlink of the heterogeneous network. 
The ultimate goal is to maximize the long-term utility of the network while ensuring QoS requirements. 

In \cite{nasir2018deep}, the authors use a multi-agent DQL approach to allocate power in wireless networks. The principal objective is to maximize the weighted sum-rate of the system. 
In \cite{meng2019power}, the authors propose different DRL architectures such as {\em REINFORCE}, DQL and deep deterministic policy gradient (DDPG) for power allocation in multi-user cellular networks. The ultimate target is to maximize the overall sum-rate of the network.
In \cite{naparstek2019deep}, the authors use the DRL approach for dynamic spectrum access in wireless networks. 
The primary goal is to maximize each user's specific network utility in a distributed manner, i.e. without exchanging information.

Most of the related existing studies focus on power allocation at the small base stations (SBSs) or cognitive radios (CRs) on a small-scale network set up and take a completely distributed DRL approach (e.g. the SBSs or the CRs do not cooperate or exchange information among themselves). Some of the studies perform power allocation on multi-cell wireless networks on a large scale but in a distributed manner. Therefore, the solutions can be very far from the optimal solution. Also, these studies either consider one user with multiple subbands per cell or multi-user with one shared frequency band per cell. Therefore, these approaches do not apply to multi-cell networks with multiple users per cell sharing the same frequency subbands. 

\section{System Model} \label{system}
\subsection{Problem Formulation}
We consider a downlink  cellular network of $K$ base stations (BSs). Each BS $k \in \{1, \cdots, K\}$ has  $F$ frequency sub-bands. The bandwidth of each sub-band is $B$ MHz. The power allocated by cell $k$ in frequency sub-band $f$ is ${P}_{k,f}$ which is discrete. The total power of a cell $k$ is limited by a maximum value $P_{k}^{\max }$ such that $ \sum _{f\in F} P_{k,f} \leq P_{k}^{\max },\quad \forall k\in \{1, \cdots, K\}$.
Let $\mathcal {U}_{k}$ denote the set of users who are associated with cell $k$, and $\mathcal{U}$ is the set of all users in the network. The vector $\mathcal{A}_{k,f}$ denotes allocation of sub-band in cell $k$, where each element ${A}_{k,f}$ is an integer denoting the user who is assigned sub-band $f$ in cell $k$. 

The corresponding throughput maximization problem is given by 

\begin{align}
    & {\rm max} \sum _{k\in \{1, \cdots, K\}}\sum _{u\in \mathcal {U} _{k}} \sum _{f=1}^{F} \left [{ \mathbb{I}({A}_{k,f}=u)  B \log \left ({1 + \alpha {\textrm {SINR}}_{u,k,f}}\right )}\right ] \\
            & {\rm s.t.}  \sum _{f\in F} P_{k,f} \leq P_{k}^{\max },\quad \forall k\in \{1, \cdots, K\}
\end{align}

where  $\alpha $ is a constant for a given target Bit Error Rate (BER) which is defined as $\alpha = -1.5 / \log (5\textrm{BER}) $. We assume $\textrm{BER}$ to be $10^{-6}$. The signal-to-interference-plus-noise ratio (SINR) of user $u$ when served by cell $k$ which transmits over frequency sub-band $f$ is expressed as
$ {\textrm {SINR}}_{u,k,f} = \frac {P_{k,f} G_{u,k,f}} {\eta _{u} +\sum _{l \neq k} P_{l,f} G_{u,l,f}}$.
where $\eta _{u}$ represents the receiver noise and $G_{u,k,f}$ denotes the link gain from cell $k$ to user $u$ over frequency sub-band $f$  defined as 
$
    G_{u,k,f}=10^{-(PL_{u}+X_{\alpha})/10}.|H_{u,k,f}|^2,
$
where $H_{u,k,f}$ is the Rayleigh fading gain of user $u$ from cell $k$ over frequency sub-band $f$, $X_{\alpha}$ is the log-normal shadowing, and $PL_{u}$ is the path-loss of user $u$. 

The utility of the network which is the total network throughput is defined as follows: 
\begin{equation}\label{U}
U =\sum _{k\in \{1,\cdots,K\}}\sum _{u\in \mathcal {U} _{k}} \sum _{f=1}^{F} 
\left [{ \mathbb{I}({A}_{k,f}=u)  B \log \left ({1 + \alpha {\textrm {SINR}}_{u,k,f}}\right )}\right].
\end{equation}

All users periodically send their channel quality as a channel quality indicator (CQI) to their nearest BS, where
$\mathcal{C}_{u,k}$ is the CQI vector of a user $u$ of cell $k$ over all frequency sub-bands. That is, $\mathcal{C}_{u,k}$ is a  vector of discrete values   $\mathcal{C}_{u,k} \mathrel {\mathrel {\mathop :}\mkern -1.2mu=}(\mathcal{C}_{u,k,1}, \mathcal{C}_{u,k,2},\ldots, \mathcal{C}_{u,k,F})$. For example, in LTE, the CQI value ranges from $1$ to $15$ \cite{cqi}. In addition, users are also classified into cell-center users and cell-edge users depending on the users' locations. A location indicator $\mathcal {V}_{u,k}$ is used to indicate whether a user $u$ of cell $k$ is cell-center user or cell-edge user as follows:
\begin{equation} 
\label{eq_location}
\mathcal {V}_{u,k} = 
\begin{cases}
1 & \mbox {if}~~{{R}_{u,k}> R/2} \\
0 & \mbox {otherwise} \end{cases} 
\end{equation}
where ${R}_{u,k}$ is the distance of user $u$ in cell $k$ from the BS and $R$ is the cell radius. Subband $f$ of cell $k$ will be allocated to an user $u\in \mathcal {U} _{k}$ for which it will maximize the throughput of subband $f$ in cell $k$. Therefore, the subbands are allocated based on the following equation:  

\begin{equation} 
\label{eq_channel}
A_{k,f}= {\arg\max}_{u\in \mathcal{U}_{k}} B \log \left ({1 + \alpha {\textrm {SINR}}_{u,k,f}}\right )
\end{equation}

\section{Power Allocation in Multi-cell Networks: A DRL Approach}\label{proposed}
In the following, we develop a DRL-based resource allocation model for multi-cell networks with multiple users per cell sharing the same frequency subbands with an objective of maximizing the total network throughput by performing power allocation. 

\subsection{Basics of DRL}
\begin{figure}[!]    
\centering 
\includegraphics[width=\linewidth]{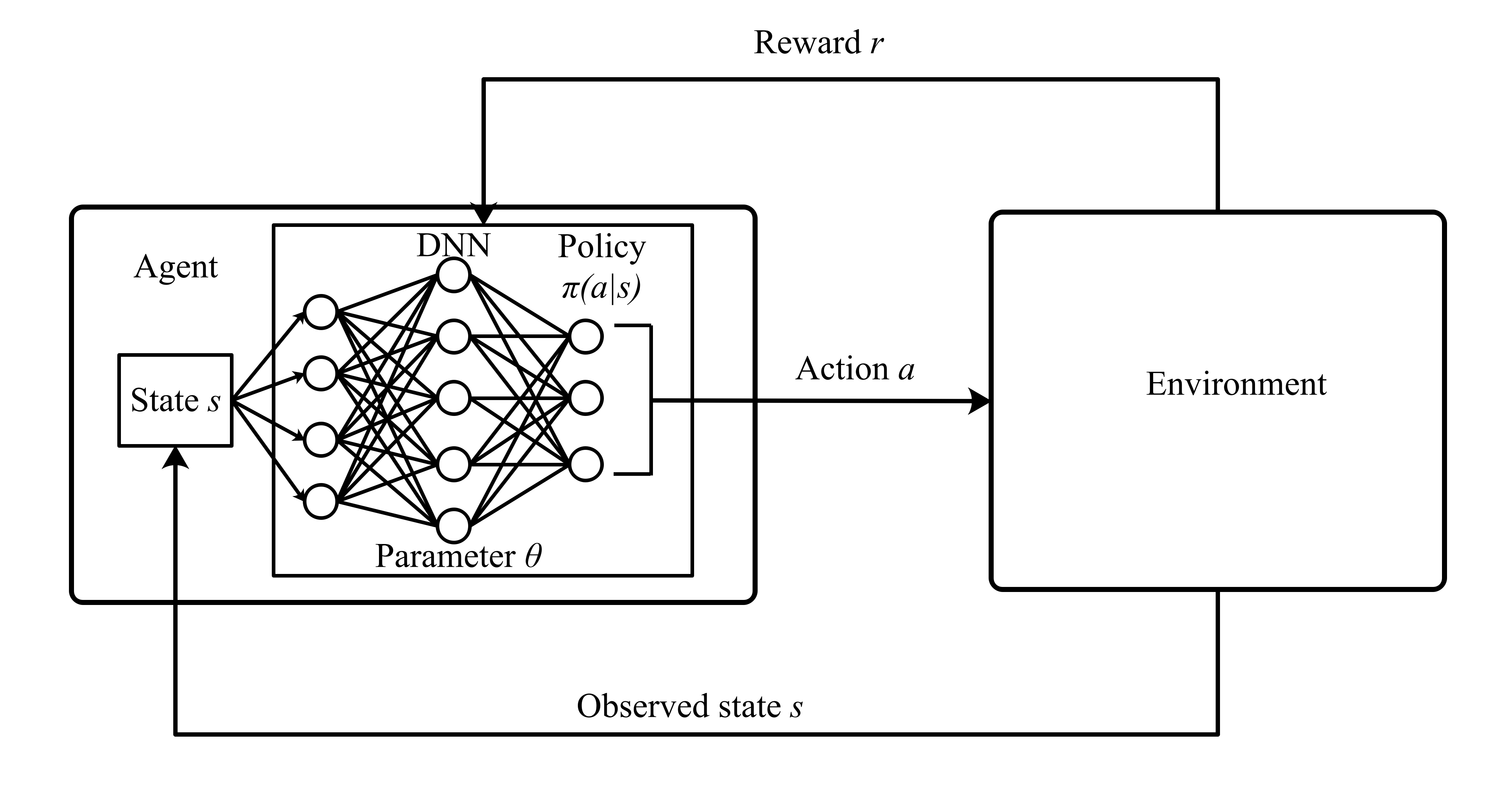}
\caption{Deep reinforcement learning.}
\label{fig_deep1}
\end{figure}

DRL is a combination of deep neural network (DNN) and reinforcement learning (RL). In DRL, a DNN works as a software agent and interact with the environment as shown in in Fig.~\ref{fig_deep1}.
In this article, we focus on the DQL. At time step $t$, the DQL agent receives state $s_t$ from a state space $\mathcal{S}$ and takes an action $a_t$ from an action space $\mathbb{A}$. The agent follows a policy $\pi(a_t|s_t)$ i.e., a mapping from state $s_t$ to action $a_t$, to choose the action. After executing action $a_t$, the agent receives a reward $r_t$ and move to new state $s_{t+1}$. The agent continues the process until it reaches the terminal state and then it restarts. The goal of the agent is to maximize the discounted accumulated reward defined as $R_t = \sum_{k=0}^{\infty} \gamma^k r_{t+k}$. Here, $\gamma \in (0,1]$ is the discount factor which determines the importance of future rewards compared to current reward. An action-value function $Q_{\pi}(s,a)=E[R_t|s_t=s, a_t=a]$ is the expected return for selecting action $a$ in state $s$ and then  follow a policy $\pi$. An optimal action-value function $Q^{*}(s,a)=\max_{\pi}Q_{\pi}(s,a)$ is the maximum action value achievable by following any policy for state $s$ and action $a$. The optimal action-value function can be expressed by the Bellman equation as follows: 

\begin{equation}\label{Q*}
Q^{*}(s,a)= \mathbb{E}_{s^{'}}[r+\gamma \max_{a^{'}}Q^{*}(s^{'},a^{'})|s,a].
\end{equation}

In DQL, we use a neural network to approximate the optimal action-value function, $Q(s,a;\theta)\approx Q^{*}(s,a)$. Here, $Q(s,a;\theta)$ is called the Deep Q-network (DQN) and $\theta$ is the parameter of the neural network. The iterative update is used to train the Q-network and thus reduce the mean-squared error of the Bellman equation. 

\subsection{DQL Approach}
Now we present a DQL approach that can perform near-optimal power allocation on multi-cell networks. Specifically, this DQL model uses $\mathcal{C}_{u,k}$ vector along with $\mathcal {V}_{u,k}$ of all users in a network as state and then takes action. Here, each action corresponds to a power allocation. That is, for $K$ cells, $U$ users and $F$ sub-bands,  the state size is $(K \times U\times (F+1))$. The total number of actions depends on the number of power levels we are using. For $n$ number of power levels and $F$ frequency sub-bands, we can have a maximum $n^F $ power combinations. Some of the combinations will be discarded due to the maximal power constraint. Let $m$ denote the total number of combinations possible where each combination corresponds to an action. For each cell, we have $m$ number of actions. Let $\mathbb{A}_k$ denote the action space for $k$ cell and $\mathfrak{a}_k \in \mathbb{A}_k$ is the selected action for cell $k$. Therefore, for $K$ number of cells, we have $K\times m$ number of actions. The DQN model has to take action from $m$ number of actions for each cell. Therefore, in total, the DQN model has to take $K$ number of actions. At time step $t$, the selected action $a_t=[\mathfrak{a}_{1},\mathfrak{a}_{2},\cdots, \mathfrak{a}_K]$. The approach proceeds in the following phases:

\subsubsection{Problem formulation}First we need to formulate the problem to apply the DQL approach. The job of the agent is to maximize the total network throughput (Eq.~(\ref{U})). The episode starts from a initial state and continues as long as the throughput increases, i.e. ${current\_throughput}> {previous\_throughput} $. Here, ${current\_throughput}$ the network throughput achieved by executing the recent actions and ${previous\_throughput}$ is due to the previous actions. The episode ends when it reaches the terminal state, i.e., the throughput decreases due to the recent actions.

\subsubsection{Training Phase} The training process of the proposed DQL based power allocation is shown in \textbf{Algorithm~\ref{alg1}}. We use DQL with experience replay \cite{mnih2015human} to train our model. In our model, the specific steps are as follows:

\noindent\textbf{Step 1:} Define the Q-Network, i.e. the number of layers and neurons per layer and the activation functions. We use the input layer size same as the state size and output layer size as the total number of actions.  We initialize two Q-network with random weights: one is for the \textbf{action-value function} $Q$ and another for \textbf{target action-value function} $\hat{Q}$. We also initialize the replay memory $D$ to some capacity $N$. 

\noindent\textbf{Step 2:} Allocate a random power vector for each cell. After that, we calculate the CQI value of each sub-band for each user in the network. We also estimate the location indicator for every user using Eq.~(\ref{eq_location}). Note that the CQI values and the location indicator for every user represent the initial state $s_t$. 

\noindent\textbf{Step 3:} Select the action which is consist of minimum power value possible for all the subbands for each cell. Then, we execute these actions and calculate the total network throughput using Eq.~(\ref{U}). Use this throughput as ${previous\_throughput} $ for next step.

\noindent\textbf{Step 4:} Use the $\varepsilon$-greedy policy to select the actions randomly or use the Q-network to choose the actions. We use the state $s_t$ as input to the Q-network to calculate the action-value for each action as we have $m$ number actions for each cell. Therefore, for the first cell, select the action which action-value is maximum among the first $m$ actions. Then, for the remaining cells, select the action consecutively with maximum action-value from the next $m$ actions and so on. 

\noindent\textbf{Step 5:} Execute the selected actions, i.e. map the actions with their corresponding power vectors and calculate the new state $s_{t+1}$ in the same way mentioned in \textbf{Step 2}. The agent then receives a positive reward of $r_t= +1$  We also calculate the total network throughput using Eq.~(\ref{U}) and save the value as ${current\_throughput} $. Then, we check whether the new step is a terminal state or not using the following condition: ${current\_throughput}> {previous\_throughput} $. Finally, we store the transitions $(s_t,a_t,r_t,s_{t+1})$ in replay memory $D$. 

\noindent\textbf{Step 6:} Perform experience replay on the Q-network. The experience replay mechanism has the following steps.

\noindent\textbf{Step 6.1:} Sample a minibatch of transitions of size $\mathcal{M}$ randomly from the replay memory $D$.

\noindent\textbf{Step 6.2:} Update the targets of that minibatch. We use the target action-value $\hat{Q}$ network to generate the targets. 

\noindent\textbf{Step 6.3:} Perform a gradient descent step on the loss function to update the action-value Q-network parameters. 

\noindent\textbf{Step 6.4:} Clone the action-value function $Q$ parameters to get the target action-value function $\hat{Q}$ at every $B$ updates. 

\noindent\textbf{Step 7:} Repeat \textbf{Steps 5-6} until the agent reaches the terminal state. 

\noindent\textbf{Step 8:} Repeat \textbf{Steps 2-7} to train the Q-network for certain amount of time. 

\subsubsection{Testing Phase} After training our model, we need to test how close the model can predict compared to the optimal one in terms of total network throughput. 
One way to find the optimal solution is to check all possible combinations of power allocation for all the BSs which is referred to as {\em exhaustive search}. For example, with $15$ cells, $5$ sub-bands, and $5$ discrete power levels, then there will be $5^5$ or $ 3125$ possible combinations available for the power setting of one BS. Therefore, we will need to check $3125^{15}$ combinations, which is practically infeasible.  Therefore, we resort to a GA~\cite{sivanandam2008genetic}, which is a heuristic searching algorithm inspired by the theory of natural evolution. 

For testing, the following steps are added to the training steps:

\noindent\textbf{Step 9:} Repeat \textbf{Steps 2-7} and save the second last action and the network throughput for that action. Therefore, if $s_{t+1}$ is the terminal state and $a_t$ is the action for which the agent reaches the terminal state, then we need to save the action $a_{t-1}$ and the network throughput for that action. Here, the action $a_{t-1}$ is considered to be the optimal action.  

\noindent\textbf{Step 10:} Find the power vector of every BS which maximizes the total network utility (Eq.~(\ref{U})).  We use GA to solve this problem.

\noindent\textbf{Step 11:} Apply the optimal power vector solution to Eq.~\ref{U} to calculate the total network throughput. We also save the optimal solution and network throughput. 

\noindent\textbf{Step 12:} Keep repeating the steps until we have a certain amount of testing data.  

All the training and testing will be performed online. In a practical setting, all users in the network periodically send their CQI values to their serving BSs, which extract the CQI value of each sub-band and add a location indicator, i.e. cell-centre user or cell-edge user. Therefore, for every user, there will be a vector of CQI and location indicator. Each BS then sends the processed information of all users to a central entity (e.g. SDN controller), which runs the DQL model. The DQL agent selects the power vector for all the BSs as an action. Once the agent chooses the action, the controller will send back the power vectors to their designated BSs. The BSs then allocate the power accordingly. 

\begin{algorithm}
    \caption{DQL with experience replay for power allocation}
    \label{alg1}
\begin{algorithmic}[1]
 \renewcommand{\algorithmicrequire}{\textbf{Input:}}
 \renewcommand{\algorithmicensure}{\textbf{Output:}}
    \STATE Initialize replay memory $D$ to capacity $N$
    \STATE Initialize action-value function $Q$ with random weights $\theta$
    \STATE Initialize target action-value function $\hat{Q}$ with weights $\theta^{-}=\theta$
    
    \FOR {$\mbox{episode} = 1,\cdots,M$}
    \STATE Allocate a random power vector for each cell.
    \FOR {$t = 1, \cdots, \infty$}
    \STATE Calculate the CQI vector as well as the location indicator for every user in the network.
    \STATE Use the CQI vector and the location indicator as state $s_t$.
    \FOR {$\mbox{k} = 1, \cdots, K$}
    \STATE With probability $\varepsilon$ select a random action $\mathfrak{a}_k$ for cell $k$
    \STATE Otherwise select $\mathfrak{a}_k={\arg\max }_{a \in \mathbb{A}_k}Q(s_t,a;\theta) $
    \ENDFOR
    \STATE Execute action $a_t=[\mathfrak{a}_{1},\mathfrak{a}_{2},\cdots, \mathfrak{a}_K]$ and observe reward $r_t$ and state $s_{t+1}$
    \STATE Store transition $(s_t,a_t,r_t,s_{t+1})$ in $D$
    \STATE Sample random minibatch of transitions $(s_j,a_j,r_j,s_{j+1})$ from $D$
    \STATE Set ${y}_{j} = r_j$ if episode terminates at step $j+1$
    \STATE Otherwise set $y_j = r_j+\gamma \max_{a^{'}} \hat{Q}(s_{j+1},a^{'};\theta^{-})$
    \STATE Perform gradient descent step on ${(y_j-Q(s_j,a_j;\theta))}^2$ with respect to the network parameters $\theta$
    \STATE Every $B$ steps reset $\hat{Q}=Q$
    \ENDFOR
    \ENDFOR
\end{algorithmic} 
\end{algorithm}

\section{Experimental Setup}\label{result}
We present the simulation settings of the proposed DQL based power allocation scheme. We implement the proposed algorithm using Tensorflow \cite{45381}. We consider three different simulation scenarios: Scenario $1$ with $K$ = $5$ BSs, Scenario $2$ with $K$ = $10$ BSs, Scenario $3$ with $K$ = $15$ BSs, cell coverage radius $R=500$m, maximum transmit power = $40$W, directional antenna per cell = $3$, number of users per cell = $5$, bandwidth of a sub-band $B$ = 2.88MHz, white noise power density = $-174$dBm/Hz, number of sub-bands = $3$, number of power levels = $5$, and power levels = $\{ 6.4, 9.6, 12.8, 16, 19.2\}$W. 
\begin{figure*}[!h] 
    \centering
  \subfloat[Scenario $1$: $5$ cells\label{1a}]{%
       \includegraphics[width=0.33\linewidth]{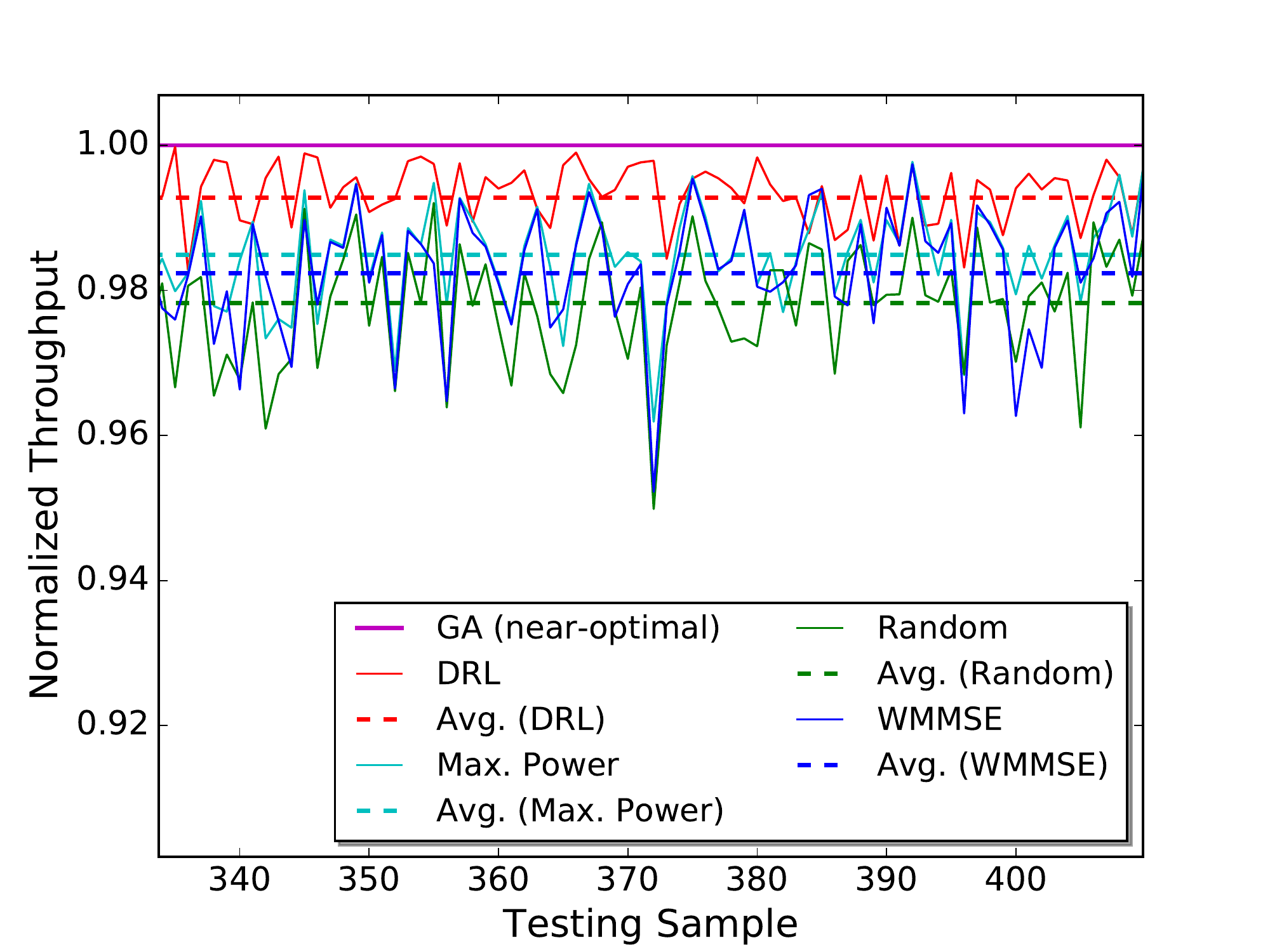}}
  \subfloat[Scenario $2$: $10$ cells\label{1b}]{%
        \includegraphics[width=0.33\linewidth]{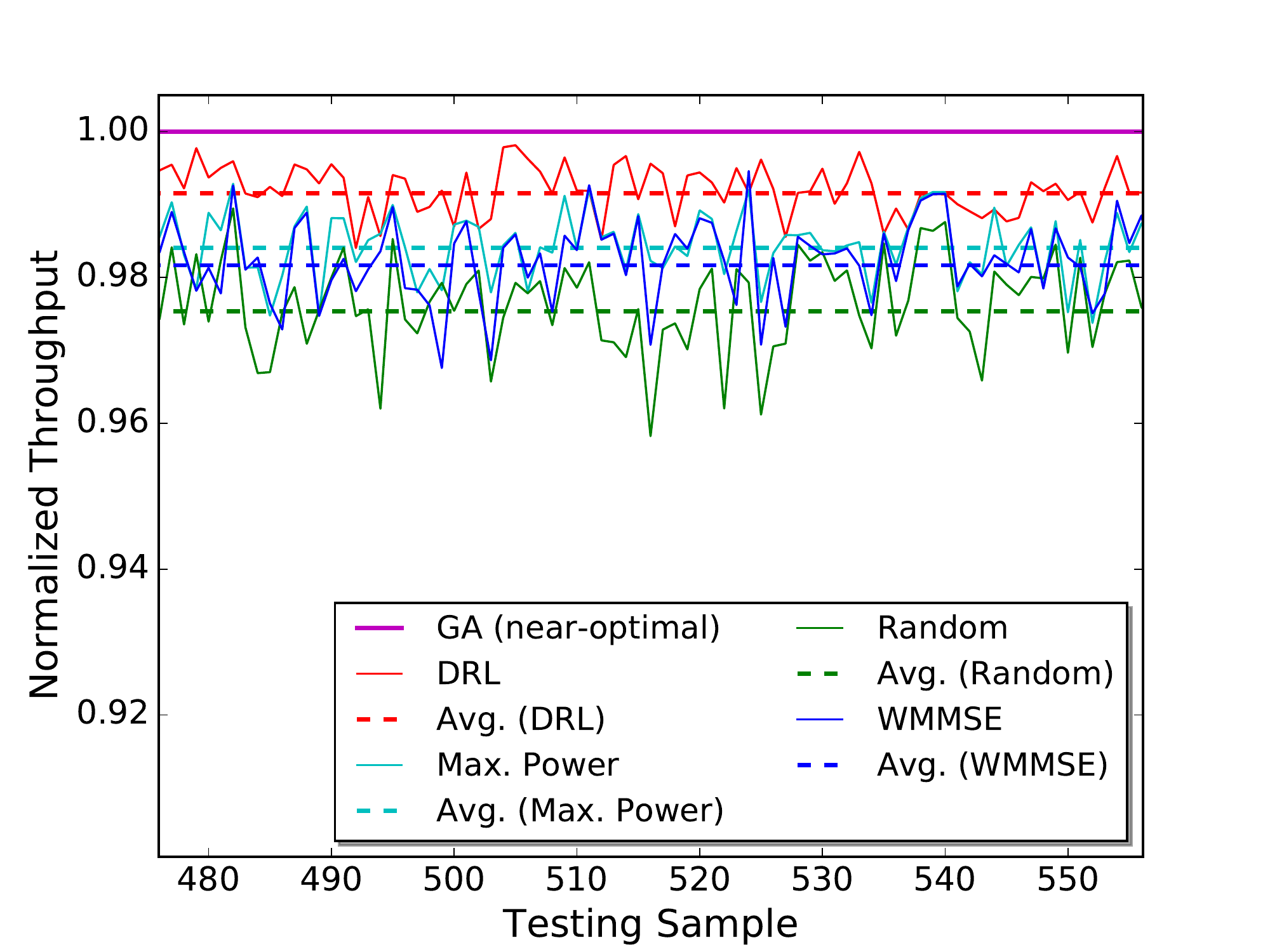}}
  \subfloat[Scenario $3$: $15$ cells\label{1c}]{%
\includegraphics[width=0.33\linewidth]{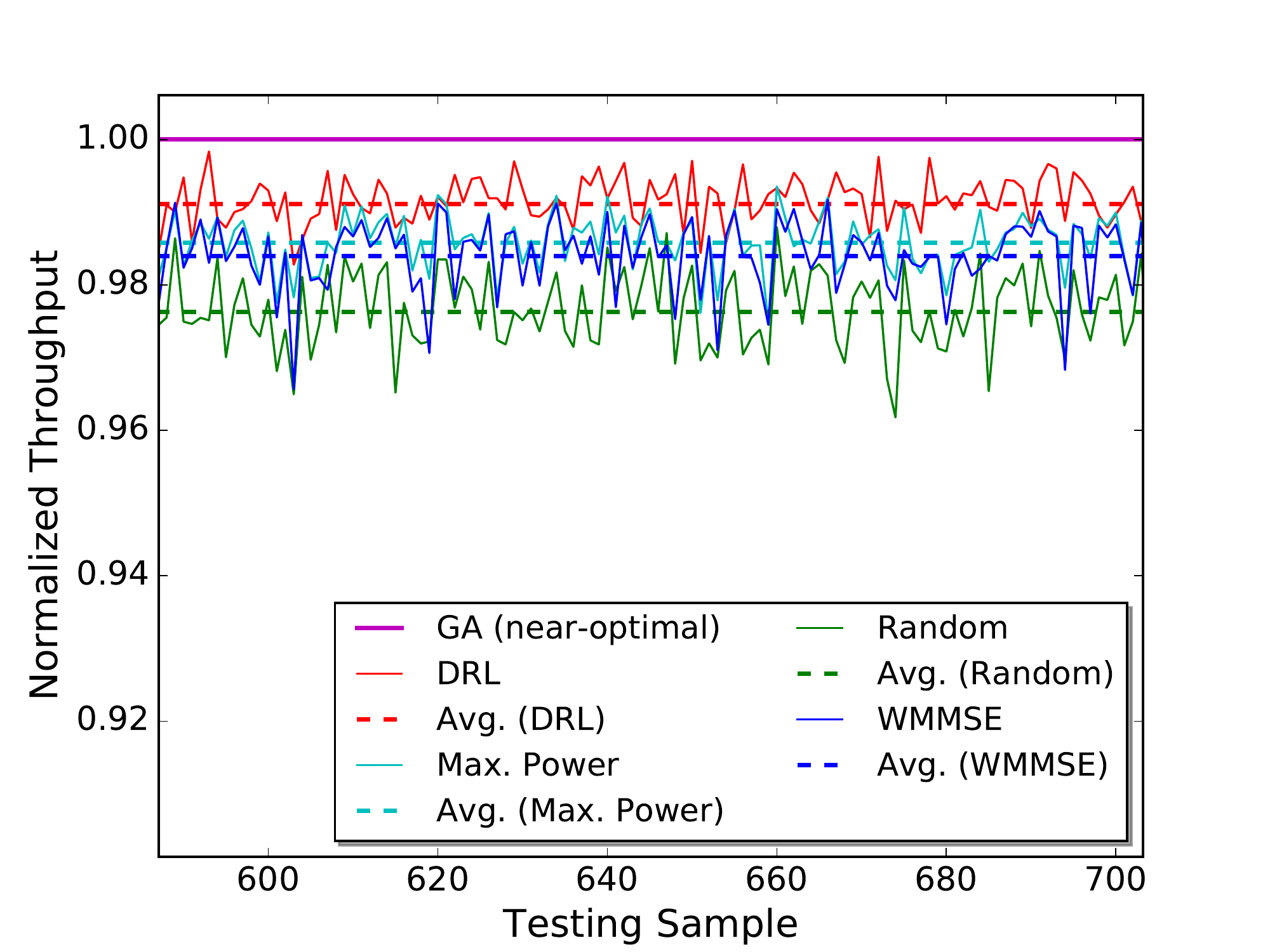}}
\caption{Normalized throughput vs. testing samples. }
  \label{fig2_1} 
\end{figure*}
\subsection{Training the DQL model}
We first need to define the DQN. We use a deep neural network of one hidden layer as our DQN. We use Rectified Linear Unit (ReLU) as an activation function for the hidden layer. The state size, i.e. the input layer size of the Q-network for three different scenario is $5\times 5\times (3+1)=100$, $10\times 5\times (3+1)=200$, $15\times 5\times (3+1)=300$. For each cell, the total number of power combinations possible for $5$ power levels is $5^3=125$. Some of the power combinations will be discarded due to the maximal power constraint. After applying the limitation, we have $72$ power combinations for each cell. So, the total number of actions, i.e. the output layer size of the DQN for three different scenario is $5\times 72=360$, $10\times 72=720$, $15\times 72=1,080$. The training parameters of the DQN are shown in Table~\ref{table:1}.

\begin{table}[H]
\centering
\caption{Training parameters}
\label{table:1}
\begin{tabular}{|l|l|lll}
\cline{1-2}
\textbf{Parameter}& \textbf{Value}  &  &  &  \\ \cline{1-2}
Number of hidden layers & $1$  &  &  &  \\ \cline{1-2}
Layers & \begin{minipage}{35mm} \{Input, Hidden Layer, Output\} \end{minipage}  &  &  \\ \cline{1-2}
No. of neurons per layer &  \begin{tabular}[t]{@{}l@{}l@{}}Scenario $1: $ $\{100, 720, 360\}$ \\Scenario $2: $  $\{200, 1440, 720\}$ \\Scenario $3: $  $\{300, 2160, 1080\}$ \end{tabular}  &  &  &  \\ \cline{1-2}
Replay memory size  & $80,000$  &  &  &  \\ \cline{1-2}
Batch size & 64  &  &  &  \\ \cline{1-2}
Update target frequency, $B$ & $1000$  &  &  &  \\ \cline{1-2}
Learning rate & $0.00025$  &  &  &  \\ \cline{1-2}
Loss function & MSE  &  &  &  \\ \cline{1-2}
Optimizer & RMSprop  &  &  &  \\ \cline{1-2}
No. of epochs per training & $1$  &  &  &  \\ \cline{1-2}
\end{tabular}
\end{table}

\subsection{Testing the DQL model} 

After training the DQL model for about $80 ~\mbox{hrs}$, we compare our model with the optimal power allocation. We use the GA approach to find near-optimal power allocation. We also compare our power allocation (PA) model with other power allocation models such as WMMSE \cite{shi2011iteratively}, maximum power allocation (PA), and random power allocation model. We use $12.8$ W power for each subband for maximum power allocation. 

\subsection{Results and Discussions}
For comparison purpose, we calculate the total network throughput for the power allocation solution achieved through different PA model as well as for the near-optimal PA solution derived through GA. Then we calculate the normalized total network throughput of different PA models by dividing it with the total network throughput of the GA solution. Fig.~\ref{fig2_1} shows the normalized throughput of different PA models vs. testing samples for different network scenarios. 
From the figure, it is evident that the proposed DRL-based PA model performs better than other PA models.


\subsubsection{Impact of the wireless network size on the DRL performance}
The average normalized throughput from our proposed DRL based PA model for scenario-$1$ is $0.99276$,  scenario-$2$ is $0.99157$ and scenario-$3$ is $0.99109$.  From Fig.~\ref{fig2_1}, it is also evident that with the increase of the wireless network size (i.e. number of cells), the performance of the proposed DRL-based PA model decreases gradually. The average normalized network throughput decreases with the increase of the wireless network size. This is because, with the increase of the wireless network size, the state space and the action space also increases. As a result, the DQN needs to explore more state-action space to find the optimal action policy. Therefore, more exploration is required for large state-action space. This is why the performance of our DRL model degrades gradually with the increase of wireless network size.  

\subsubsection{Impact of the hidden layer size of the DQN on the DRL performance}
\begin{figure}[!htb]
\minipage{0.4\textwidth}
  \includegraphics[width=\linewidth]{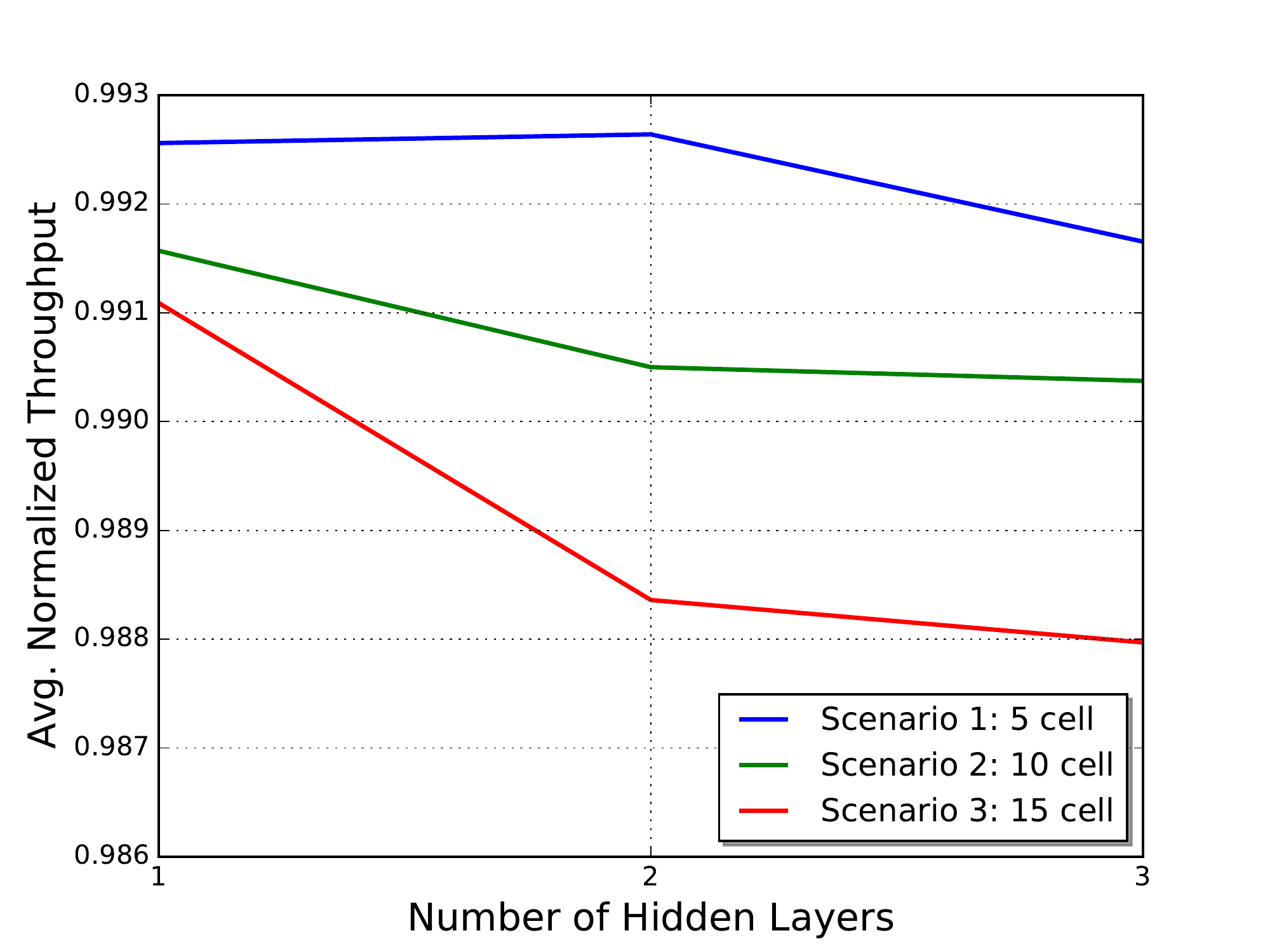}
  \caption{Average normalized throughput vs. number of hidden layers.}\label{fig3}
\endminipage\hfill
\minipage{0.4\textwidth}
  \includegraphics[width=\linewidth]{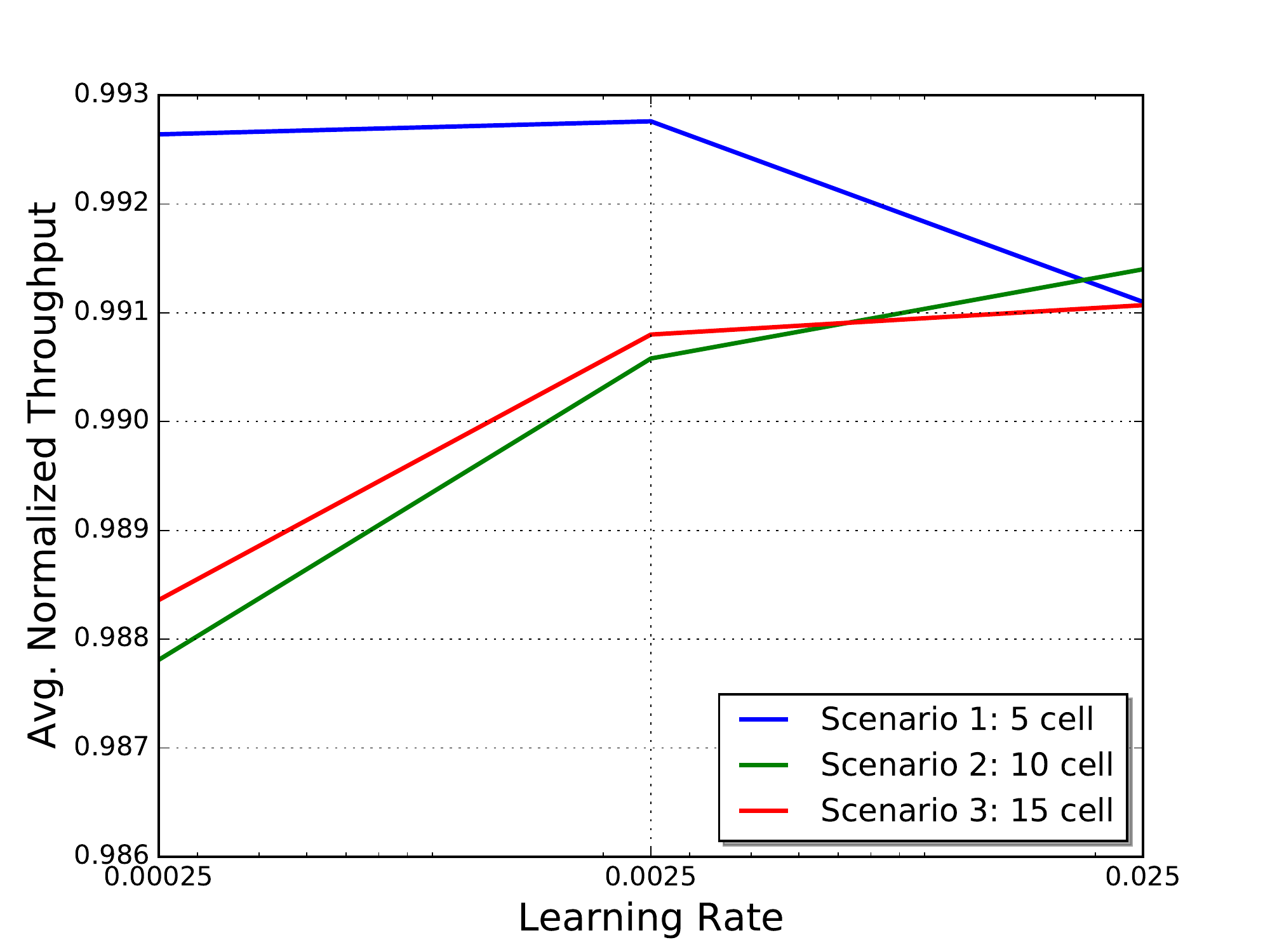}
  \caption{Average normalized throughput vs. learning rate.}\label{fig4}
\endminipage\hfill
\end{figure}

The number of hidden layers of the DQN is an important parameter as the DQN approximate the action-value function ($Q$). The DQN approximates the state-action relationship by extracting useful information from the sate. More hidden layers in DQN means it can learn more features. We vary the hidden layer size of the DQN and repeat the simulations. Fig.~\ref{fig3} shows the average normalized throughput of the proposed model vs. the number of hidden layers for different network scenarios. It is apparent that the performance of the DRL model slightly degrades with the increase of the hidden layer size of the DQN. This is because with the increase of more hidden layers, the DQN learns irrelevant features (noise) and as a result of that {\em{overfitting}} occurs which eventually degrades the performance of the DQN.   

\subsubsection{Impact of the learning rate on the DRL performance}
The learning rate is an important hyperparameter that controls the amount of change in weights of the DQN during the training procedure. It controls how quickly or slowly a DQN learns from data. Finding the optimal learning rate is challenging since a small learning rate may result in larger training time and a large learning rate may result in an unstable training process. Next, we vary the learning the rate of the DRL keeping other parameters fixed. Fig.~\ref{fig4} shows the average normalized throughput of the proposed model vs. learning rate for different network scenarios. The optimal learning rate for scenario-$1$ is $0.0025$ and for scenario-$2$ and scenario-$3$ is $0.025$. 

\section{Conclusion}\label{concl}
We have presented a novel DRL-based method for power allocation in multi-cell networks. Specifically, we have used DQL with experience replay for the proposed method. Simulation results show that the DQN with one hidden layer is enough to approximate the action-value function for our case. The learning rate is the most important hyperparameter in DRL and finding optimal learning is challenging. To find the optimal learning rate for different network scenarios, we have varied the learning rate and observed the performance of the proposed model. We have evaluated the performance of the proposed DRL-based power allocation method with other power allocation methods such as WMMSE, maximum power allocation, and random power allocation for different network scenarios. Simulation results show that the proposed method is scalable to large-scale scenarios and it performs better compared to other PA models.

\bibliography{IEEEfull,references}
\bibliographystyle{IEEEtran}

\end{document}